\documentclass[12pt]{spieman}  
\usepackage{amsmath,amsfonts,amssymb}
\usepackage{graphicx}
\usepackage{setspace}
\usepackage{tocloft}

\title{Image steganography based on generative implicit neural representation}

\author[a]{Yangjie Zhong}
\author[a]{Yan Ke}
\author[a]{Meiqi Liu}
\author[a,b,*]{Jia Liu}
\affil[a]{Engineering University of PAP, Shanxi, China, 710000}
\affil[b]{Key Laboratory of Network and Information Security of PAP, Shanxi, China, 710000}

\cftpagenumbersoff{figure}
\cftpagenumbersoff{table} 
\begin{document} 
\maketitle

\begin{abstract}
In the realm of advanced steganography, the scale of the model typically correlates directly with the resolution of the fundamental grid, necessitating the training of a distinct neural network for message extraction. This paper proposes an image steganography based on generative implicit neural representation. This approach transcends the constraints of image resolution by portraying data as continuous functional expressions. Notably, this method permits the utilization of a diverse array of multimedia data as cover images, thereby broadening the spectrum of potential carriers. Additionally, by fixing a neural network as the message extractor, we effectively redirect the training burden to the image itself, resulting in both a reduction in computational overhead and an enhancement in steganographic speed. This approach also circumvents potential transmission challenges associated with the message extractor. Experimental findings reveal that this methodology achieves a commendable optimization efficiency, achieving a completion time of just 3 seconds for 64x64 dimensional images, while concealing only 1 bpp of information. Furthermore, the accuracy of message extraction attains an impressive mark of 100\%. 
\end{abstract}


{\noindent \footnotesize\textbf{*}Jia Liu,  \linkable{liujia1022@gmail.com} }

\begin{spacing}{2}   

\section{Introduction}
Steganography represents a communication technology that leverages public channels for covert information transmission\cite{fridrich2009steganography}. A diverse array of media, including images\cite{tao2018towards}, digital audio\cite{yi2019ahcm}, video\cite{xu2014data}, and text\cite{borges2008robust}, serve as platforms for concealing sensitive information. Presently, generative steganography, grounded in deep learning techniques, has garnered significant research interest in the domain of steganography. Unlike traditional counterparts, generative steganography relies heavily on sophisticated machine learning methodologies and models, employing techniques such as generative adversarial networks\cite{goodfellow2014generative} and diffusion models\cite{tang2017automatic}. Its fundamental premise shifts from embedding secret information within existing content to directly crafting content that inherently incorporates secret information\cite{zhang2019steganogan}, rendering these contents visually indistinguishable from standard data.

While generative image steganography offers significant advantages in terms of information concealment, it is not without its challenges. Firstly, image resolution is inherently limited. In steganography schemes, cover images are represented as a grid and stored as pixel points in computers. However, when continuous image signals are converted into discrete grid data through sampling, crucial detail information is inevitably lost, compromising both image resolution and quality. Secondly, the training of extractors incurs substantial costs. Extractors tend to lack adaptability to varying carrier types; for instance, a specific extractor may excel in handling certain image file formats yet falter with others. Improving their performance requires extensive training, which is computationally expensive. Furthermore, the retrieval of secret information by extractors is often hampered by channel interference, such as compression, cropping, and format conversion. These factors can significantly impact extraction accuracy, leading to suboptimal results. Lastly, the transmission behavior of message extractors may arouse suspicion during steganalysis, thereby increasing the risk of exposure. Additionally, the large size of these extractors can render their transmission challenging.

\begin{figure}
\begin{center}
\begin{tabular}{c}
\includegraphics[width=\textwidth]{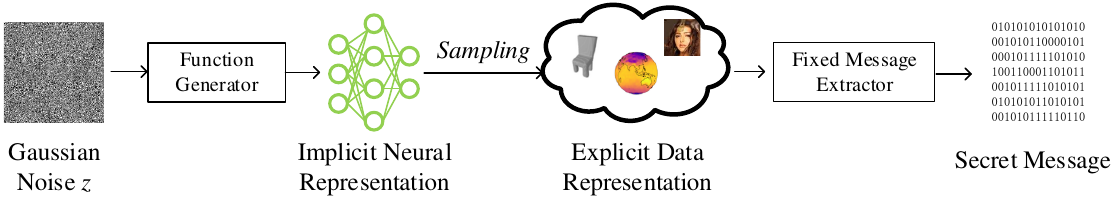}
\end{tabular}
\end{center}
\caption 
{ \label{fig:methods}
The basic idea of our scheme. } 
\end{figure} 

We propose a scheme called Image Steganography Based on Generative Implicit Neural Representation (GINR-Stega) to address the aforementioned challenges. As shown in Figure \ref{fig:methods}, firstly Gaussian noise is input, multimedia data is represented as a function using implicit neural representation through a function generator, sampling is done to get explicit data represented as a two-dimensional image, and finally, it is input into a fixed neural network for message hiding and extraction. Implicit neural representation(INR) refers to the encoding of input signals, such as images, audio, and point clouds, as functions through neural networks\cite{sitzmann2020implicit}. Our scheme incorporates generative adversarial stochastic process (GASP) and utilizes a function generation model\cite{dupont2022generative} to represent the cover image as a continuous function. This approach effectively surpasses limitations imposed by resolution and carrier type. Additionally, we employ a fixed neural network\cite{kishore2022fixed} instead of a traditional extractor, shifting the focus of training from the network to the image itself. This strategy significantly reduces extraction error rates and training costs. Our contributions are as follows:

1. This is the first instance of integrating a function generation model with a steganography scheme. Experimental results demonstrate that the function generation model exhibits versatility in data representation, suggesting promising potential for future applications in various multimedia steganography techniques.

2. Representing images as continuous functions breaks the barrier of image resolution, enabling the sampling of images with varying resolutions within low-resolution models.

3. By adopting a fixed neural network as the extractor, the extraction quality of our steganography scheme no longer hinges on the extractor's training efficacy. This approach resolves issues related to high training costs and extraction error rates, and transforms network training into image training, thereby enhancing the steganography rate.

\section{Related Work}
\subsection{Implicit neural representation}
In 2016, Ha et al.\cite{ha2016generating} groundbreakingly postulated that images can be conceptualized as functions, and further, any neural network can be viewed as a function itself. Subsequently, the notion of implicit neural representation\cite{2018Which} emerged into prominence. Park et al.\cite{Park_2019_CVPR} introduced DeepSDF, a groundbreaking approach that utilizes the concept of signed distance functions (SDF) to represent a class of shapes through functions, achieving remarkable shape representation, interpolation, and completion capabilities from incomplete and noisy 3D input data. Chen et al.\cite{chen2019learning} advocated for the utilization of implicit fields to learn generative models for shapes, introducing an innovative implicit field decoder called IM-NET to generate shapes to enhance the visual fidelity of the generated outputs. In 2020, Mildenhall et al.\cite{mescheder2020stability} proposed the Neural Radiance Field (NeRF), which achieved stunningly realistic 3D scene renderings through the lens of implicit representation. More recently, in 2022, Dupont et al.\cite{dupont2022generative} abandoned the traditional discretized grid approach, adopting continuous functions to parameterize data points. By learning the distribution within the function space, they generated data points and abstracted them as functions, transcending the confines of specific data types. This grid-independent model excels at learning rich function distributions regardless of data type or resolution, and this representation forms the foundation for the steganographic cover image generation approach presented in this paper.

\subsection{Steganography}
\subsubsection{Traditional image steganography}
Traditional image steganography encompasses three distinct approaches: cover selection, cover synthesis, and cover modification. Cover selection\cite{fridrich2009steganography}, an unadulterated information hiding technique, involves selecting a cover from a standard image library and computing its hash value based on the secret information. Cover synthesis\cite{otori2007data}, on the other hand, utilizes the cover's texture and a set of color points derived from the secret information to fuse into a dense, composite image. Cover modification, such as the Least Significant Bit (LSB) algorithm pioneered by Schyndel et al.\cite{van1994digital} and its enhanced version introduced by Mielikainen et al.\cite{mielikainen2006lsb}, conceals secret information by altering pixel data. However, as the detection capabilities of steganalysis techniques continue to advance, traditional image steganalysis methods have struggled to evade detection.

\subsubsection{Deep steganography}
Owing to the profound proficiency in feature learning exhibited by deep learning, the utilization of deep neural networks in steganography has progressively garnered significant research attention. Based on the functional delineation of deep neural networks within steganography frameworks, deep steganography can be systematically categorized into the following three distinct classes:

Steganography based on encode-decode networks. This approach harnesses the encoding and decoding capabilities of deep neural networks to process the cover image and secret message, ultimately outputting a stega-image. These methods are primarily categorized into two frameworks. The first is the Dependent Deep Hiding (DDH)\cite{baluja2019hiding} framework, where the secret information undergoes preprocessing in a network to extract its feature map. This feature map is then merged with the cover image within the coding network, resulting in a carrier that conceals the secret information. Conversely, the Universal Deep Hiding (UDH)\cite{zhang2020udh} framework focuses solely on encoding the secret information within an encoding network, independent of the cover image.Zhu\cite{zhu2018hidden} and Tancik et al.\cite{tancik2020stegastamp} have further enhanced the capabilities of codec networks in steganography. Zhu et al.\cite{zhu2018hidden} pioneeringly applied codec networks to image steganography by incorporating an analog noise processing layer. 

Steganography based on deep generation models is a prominent focus of current research. Specifically, deep generative models such as Generative Adversarial Networks (GANs)\cite{shi2018ssgan}, Diffusion Models (DPMs)\cite{ho2020denoising}, Variational Autoencoders (VAEs)\cite{kingma2013auto}, and Flow-based models\cite{kingma2018glow} are being explored to design innovative steganography algorithms.

Steganography based on GAN. In 2014, the introduction of Generative Adversarial Networks ushered in a new era of combining deep learning with information hiding. This integration can be categorized into three approaches based on the dense image generation strategy: cover modification, cover selection, and cover synthesis steganography. Methods focusing on modified steganography leverage the adversarial game strategies within GAN models to bolster steganalytic resistance. These techniques fall into two primary classes. The first group, including SGAN\cite{volkhonskiy2020steganographic} and SSGAN\cite{ho2020denoising}, employs generative models to construct an original cover image. The second class, represented by ASDL-GAN\cite{tang2017automatic} and UT-SCA-GAN\cite{yang2018spatial}, utilizes a deep generation construction modification strategy. This involves modifying the matrix with a probability of generating an adduction process while minimizing distortion costs.GANs have also been harnessed for cover-selection steganography. Ke et al.\cite{ke2019generative} pioneered a method that utilizes a set of generators to generate steganographic keys and extract messages, without the need for both parties to share datasets or maps beforehand, differing from traditional approaches. Cover synthesis steganography, based on GANs, refers to a scheme that directly generates dense vectors through a deep generation model, without specifying the original vectors beforehand. Liu et al.\cite{liu2018digital} introduced a generative steganography scheme (GSS) based on constrained sampling. Similarly, Liu et al. and Hu et al.\cite{hu2018novel} employed a message mapping approach, directly obtaining density-bearing vectors through generative models.

Steganography based on diffusion models. The diffusion model\cite{kingma2013auto,sohl2015deep} establishes a Markov chain of diffusion steps, initially introducing random noise into the data and subsequently learning the inverse diffusion process to reconstruct the desired data sample. Karras\cite{karras2022elucidating} and Xu et al.\cite{xu2023pfgm++} have capitalized on diffusion models and deterministic samplers to generate images of exceptional quality. Kim et al.\cite{kim2023diffusion} introduce a diffuse-Stego steganography scheme, which embeds secret messages into the underlying noise of a diffusion model. This innovative message projection approach is compatible with pre-trained diffusion models capable of producing high-resolution images, as well as large-scale text-to-image models.In contrast, Wei et al. propose the StegoDiffusion scheme, which eschews the traditional Markov chain approach and leverages fast sampling technology to achieve efficient steganographic image generation. This approach enables the seamless transformation between secret data and steganographic images, permitting the employment of irreversible yet more expressive network architectures.

Steganography is based on other models. Current research in steganography is not only centered on generative adversarial models and diffusion models, but also extends to schemes based on variational autoencoders (VAE)\cite{kingma2018glow} and Flow-based generative models\cite{volkhonskiy2020steganographic}. Yang\cite{yang2023flexible} introduces an innovative generative steganography framework rooted in autoregressive modeling, specifically employing PixelCNN+ for pixel-level information hiding. The framework's security is rigorously proven through theoretical derivations. Echoing the success of diffusion model steganography\cite{wei2023generative}, Jing et al.\cite{jing2021hinet} and Guan et al.\cite{guan2022deepmih} harness flow models (reversible neural networks) to devise high-capacity steganography methods. Leveraging the sensitivity of neural networks to minute perturbations, Kishore et al.\cite{kishore2022fixed} propose a steganography approach grounded in fixed neural networks. This approach revolutionizes the training process, transforming the extraction of information into the training of images, thus significantly reducing training costs.

Despite the numerous advantages of deep neural network-based steganographic techniques, including high concealment, robustness, and significant steganographic capacity, they face several noteworthy challenges. Firstly, as these networks are predominantly trained on discrete grid data, the codec's size is inherently tied to the resolution of the underlying grid. Secondly, conventional steganography schemes necessitate the transmission of message extractors to recipients, heightening the risk of exposing the steganographic activity. Thirdly, the training of message extractors is resource-intensive, often requiring significant computational power to achieve optimal extraction performance. To address these issues, Dupont et al.\cite{dupont2022generative} introduced a method that represents images using functions, thereby circumventing the resolution-based limitations. Meanwhile, Li et al.\cite{li2023steganography,li2023towards} proposed covert DNN models, which camouflage steganographic networks within ordinary deep neural networks. These covert models are designed to be integrated with image classification or segmentation tasks, ensuring the security of the message decoder. However, both these approaches still fail to mitigate the issue of high communication overhead. Kishore et al.\cite{kishore2022fixed} took a different approach, fixing the message extractor and redirecting the training focus from the network to the image itself, significantly reducing the training costs.

\subsubsection{Steganography based on INR}
In contrast to the explicit representations common in traditional steganographic approaches, neuroimplicit expressions have also found their place in modern steganographic schemes. Han et al.\cite{han2023deep} pioneered this approach by initially representing the secret message implicitly and subsequently embedding the implicit model data into a cover image. Li et al.\cite{li2023steganerf} introduced StegaNeRF, a method that integrates information within neural radiation fields (NeRF) during the rendering process. Luo et al.\cite{luo2023copyrnerf} developed an anti-distortion rendering technique by replacing the standard NeRF color representation with a watermarked version, ensuring the stability of watermark extraction from the resulting two-dimensional NeRF renderings. Chen et al.\cite{chen2023marknerf} capitalized on the unique perspective synthesis capabilities of NeRF, utilizing the secret perspective as a crucial basis for copyright verification and training a watermarking verification model through a parameterized approach. Dong et al.\cite{dong2023steganography} further extended this technology to the realm of steganography, implementing a NeRF-based steganographic scheme. All these methods leverage the radiation field as an image generator, and, akin to steganography grounded in deep neural networks, the decoder requires separate training for effective message extraction.

Based on the implicit representation of messages, Yang et al.\cite{yang2023flexible} intricately fused multiple message and carrier implicit representation models into a comprehensive network and subsequently scrambled the components. This approach represents a significant advancement in the field. On the other hand, Liu et al.\cite{liu2023hiding} introduced the groundbreaking concept of function steganography utilizing implicit representations. They achieved this by embedding the implicit representation function of the secret message directly into the carrier's implicit representation function. Remarkably, this method enables the retrieval of the secret message from the encrypted carrier solely through the exchange of a shared secret key between parties, thereby eliminating the need for a separate message extractor. Critically, this approach transforms secret message data into a unified data format—specifically, a function or model—thereby introducing a novel implementation framework for data steganography. This framework has the potential to enable steganography for a diverse range of message data types. These schemes epitomize the art of "hiding the network within the network."

Existing implicit representation steganography methods\cite{luo2023copyrnerf,chen2023marknerf,dong2023steganography,yang2023flexible} usually realize message hiding by modifying implicit functions, different from these implicit representation steganography methods, this paper applies function generation model to generative steganography for the first time, constructs the original carriers by a function generator and constructs the steganographic image by using the method of fixed neural network. Compared with the traditional generative steganography method\cite{ho2020denoising,volkhonskiy2020steganographic,tang2017automatic} that generates images directly, the scheme in this paper not only breaks through the limitation of carrier type and resolution but also solves the problem that exists in the message extractor delivery.

\section{Image Steganography Based on Generative Implicit Neural Representation}
\subsection{Framework}
\begin{figure}
\begin{center}
\begin{tabular}{c}
\includegraphics[width=\textwidth]{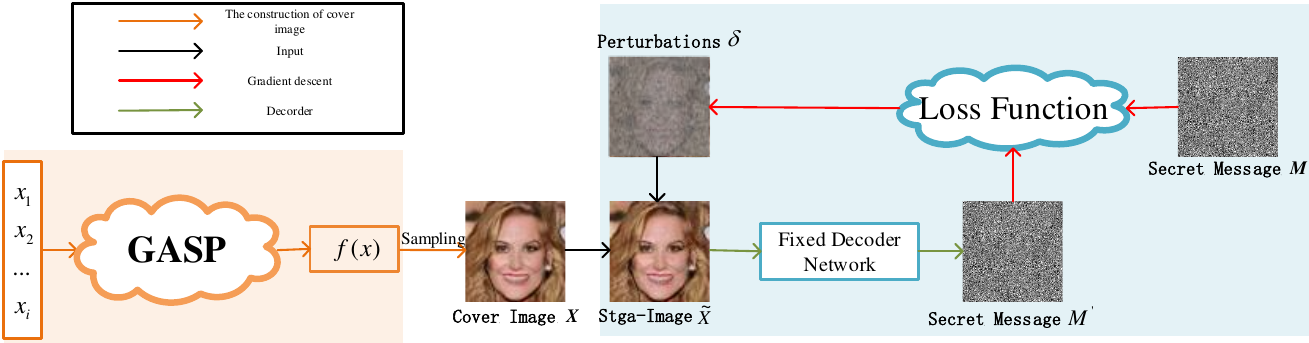}
\end{tabular}
\end{center}
\caption 
{ \label{fig:The framework of GINR-Stega}
The Framework of GINR-Stega. } 
\end{figure}

The architecture of GINR-Stega is depicted in Figure \ref{fig:The framework of GINR-Stega}. This framework encompasses the generation component of the cover image, alongside a steganographic approach leveraging a fixed neural network. To procure the original vector, we sample from the function, leveraging Dupont's generative adversarial model\cite{dupont2022generative} to obtain the desired function. Subsequently, we design the steganography scheme, building upon Kishore's fixed neural network\cite{kishore2022fixed} methodology. To acquire the cover image, we commence by sampling noise from a Gaussian distribution and inputting it into a pre-trained model that we call GASP. From this model, we sample a function, which can manifest as a two-dimensional image or diverse multimedia data, such as weather information. Then, following the desired resolution, a planar image X is sampled from the generated function. Utilizing the sensitivity of neural networks to minor perturbations, we generate a stega-image $\widetilde{X}$ from the sampled image X and a disturbed image $\delta$. For the extraction of secret information, a fixed neural network serves as a message extractor, retrieving the secret information M' from the steganographic image. Finally, we iterate between the extracted secret information M' and the original secret information M, employing a loss function to guide the modification and updating of the disturbed image. This iterative process ensures the efficacy and security of the proposed GINR-Stega-based image steganography framework.

\subsection{The Construction of Cover Image by Distribution Functions}
\subsubsection{Data representation}
In contemporary data storage practices, traditional data is typically sequestered within the confines of computer grids. However, the advent of new representation methods offers an alternative perspective. Specifically, for an image $I$, the coordinates of each pixel$(x,y)$ inherently correspond to distinct RGB feature values $\mathrm{I}(x,y)$. Although these coordinates may initially appear as a disorganized set, an underlying distance metric exists between them. Capitalizing on this intrinsic feature, the PointConv computing framework, pioneered by Wu\cite{wu2019pointconv}, is selected as a means to represent the image. The adoption of this point cloud representation for data offers significant advantages in terms of generalization.$\mathrm{x\in X}$ represents coordinates and $\mathrm{y\in Y}$ represents features, a coordinate-feature pair can be construed as a point cloud element. This approach enables the representation of any type of data, offering immense flexibility. The dataset is subsequently formulated as a comprehensive collection of coordinate-feature pairs, adhering to a point cloud format. This format is particularly advantageous due to its ambiguity in determining the data's original grid origin or the resolution at which it was sampled. Consequently, the data becomes entirely independent of the grid, operating directly on the collection itself.

\subsubsection{Implict neural expression based on distribution function}
For an image, $x=(x,y)$ corresponding to pixel positions,$y=(r,g,b)$ corresponding to RGB values, $\{(x_i,y_i)\}_{i=1}^n$ corresponding to the set of point clouds of all pixel positions and RGB values. Given a set of coordinates and their corresponding features, we can learn a function of this image $f_{\theta}\colon X\to Y$ by minimizing the function:$min_{\theta}\sum_{i=1}^{n}\|f_{\theta}(x_{i})-y_{i}\|_{2}^{2}$. The core of function representations is that they scale with signal complexity rather than signal resolution, and to represent more complex signals, we can achieve this with an extended capacity of $f_\theta$.

In generative modeling, we typically generate models based on the distribution of these data. When representing data points with a function, we want to learn the distribution of the function. In the case of images, for example, the standard generative model usually samples some noise and outputs n pixels through a neural network. In this paper, we sample the neural network weights to obtain a function that can be obtained at arbitrary coordinates. This function representation abandons the traditional underlying network and operates directly on coordinate and feature pairs. In this paper, we use an adversarial approach to train the function distribution model, called Generative Adversarial Stochastic Process\cite{dupont2022generative}. In this way, cover images can be generated and represented as continuous functions.

\subsubsection{Construction of function generator}
\begin{figure}
\begin{center}
\begin{tabular}{c}
\includegraphics[height=5.5cm]{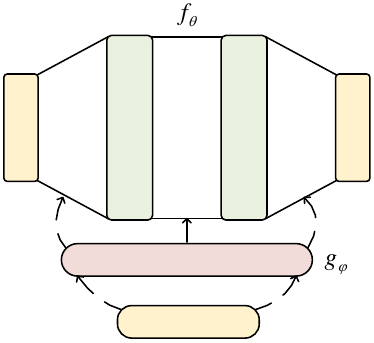}
\end{tabular}
\end{center}
\caption 
{ \label{fig:The Construction of Function Generator}
The Construction of Function Generator. } 
\end{figure}

Our goal is to learn the distribution of functions and thus generate cover images, but there is no way to know the exact functions. For images, we do not have direct access to the function that maps pixel positions to RGB values, but we do have access to a set of n coordinate-feature pairs. Such a set of coordinates and features corresponds to the input/output pairs of a function, avoiding direct manipulation of the function itself and learning the function distribution directly. The function generator architecture, shown in Figure \ref{fig:The Construction of Function Generator},$f_\theta$ is a multilayer perceptual machine, where the learned distribution of $f_\theta$ is the learned weights of $_{p(\theta)}$, which $_{p\theta)}$ are jointly determined by $g_{\varphi}$ and $p(z)$.$p(z)$ is the distribution of the hidden vectors, usually Gaussian, and $g_{\varphi}$ is the weights that map z in $p(z)$ to $\theta$ in $f_\theta$, where $g_{\varphi}{:}Z\to\Theta $. When the weights are decided, the input coordinate pairs can output the feature values, and in this way, the function is sampled from the function distribution, which is the function representation of a cover image. The image sampled in this way is not limited by the resolution and can be sampled with super-resolution according to the actual needs, getting rid of the dependence on meshing.

\subsubsection{Training of function generator}
\begin{figure}
\begin{center}
\begin{tabular}{c}
\includegraphics[width=\textwidth]{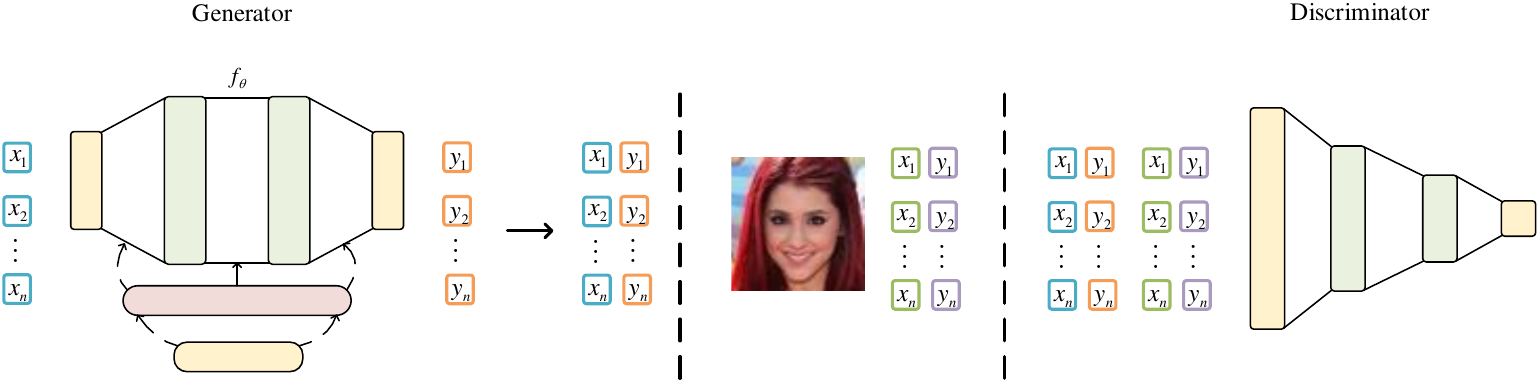}
\end{tabular}
\end{center}
\caption 
{ \label{fig:Training of Function Generator}
Training of Function Generator. } 
\end{figure}

We used Emilien's generative adversarial stochastic process\cite{dupont2022generative} to train the function generator, and the training process is shown in Figure \ref{fig:Training of Function Generator}. The construction process consists of two parts, the generator and the discriminator. Firstly, coordinate pairs are input into the generator, and a set of corresponding eigenvalues are obtained through the multi-layer perceptron, which constitutes a set of point cloud pairs. Then the real image is represented by the point cloud pair in the same way to obtain another set of coordinate eigenvalue pairs. Finally, two sets of point cloud pairs are used to judge whether the generated point cloud pairs are real through the game confrontation process. If the generated point cloud pair meets the requirements, real is returned, and the cover image is successfully constructed. Otherwise, the fake is returned. In the construction of the cover image, we refer to the experimental setting of \cite{kishore2022fixed}, use the hypernetwork parameter setting of Ha\cite{ha2016hypernetworks}, and the adversarial method of Goodfellow\cite{goodfellow2020generative}. In the training process, we do not consider using any underlying network for representation but directly sample the weight of the neural network to obtain a continuous function.

\subsubsection{Sampling}
\begin{figure}
\begin{center}
\begin{tabular}{c}
\includegraphics[height=5.5cm]{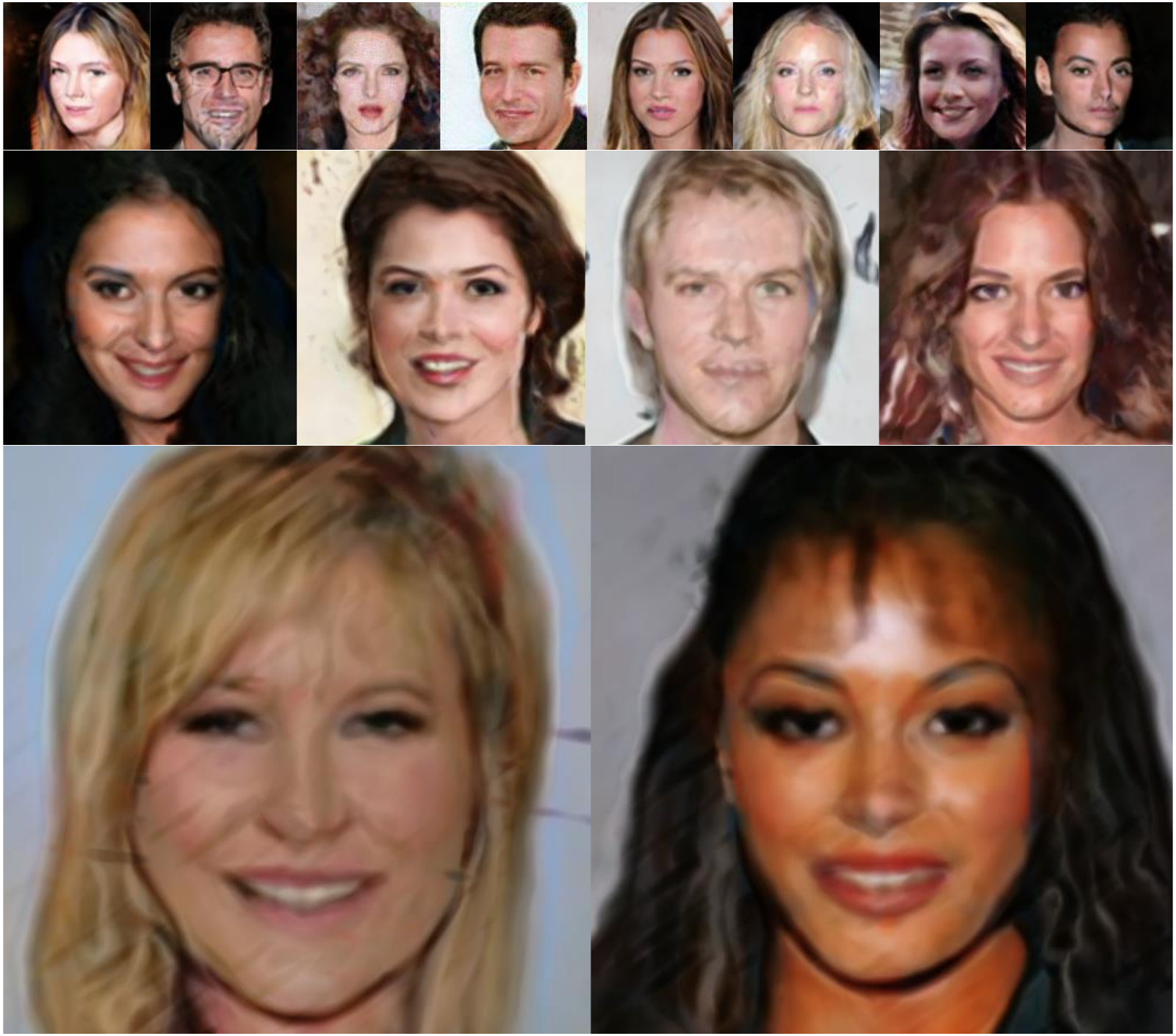}
\end{tabular}
\end{center}
\caption 
{ \label{fig:Superresolution}
Superresolution. } 
\end{figure}

The RGB value is sampled from the function by input coordinate pairs, and the function is expressed explicitly as gridded data to obtain the cover image. In this way, the obtained images are not limited by resolution. Figure \ref{fig:Superresolution} shows an example of a training data set of 64×64 dimensional images and generating high-resolution images through super-resolution sampling.

\subsection{Message Extractor Based on Fixed Neural Network}
\begin{figure}
\begin{center}
\begin{tabular}{c}
\includegraphics[width=\textwidth]{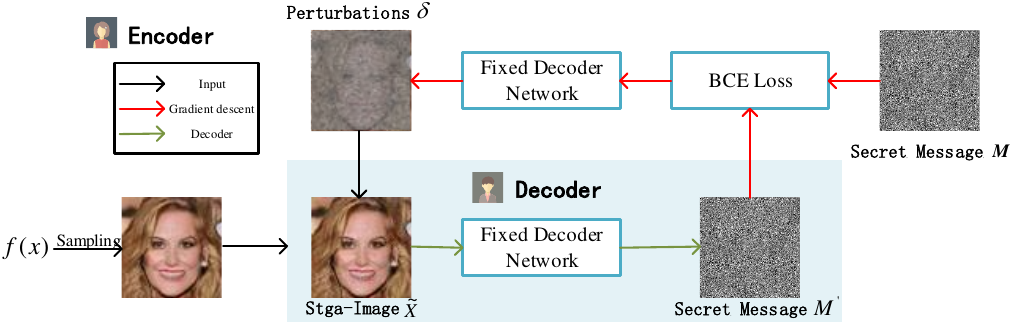}
\end{tabular}
\end{center}
\caption 
{ \label{fig:Steganography Based on Fixed Neural Network}
Steganography Based on Fixed Neural Network. } 
\end{figure}

The steganography process is shown in Figure \ref{fig:Steganography Based on Fixed Neural Network}. Firstly, a set of data is sampled from the function and expressed explicitly in the form of a two-dimensional image as a cover image, which is synthesized into a steganographic image with a disturbed image. A fixed neural network is then used as a message extractor, F is a random neural network, and any neural network with an image as an input and many binaries as an output can be used as a message extractor F. The receiver extracts the message through the fixed neural network, and the sender optimizes the extracted information and the original secret information through the binary cross-entropy loss and finally modifies the disturbed image and the cover image for an update. In this scheme, instead of training the neural network, a neural network is fixed as the message extractor. According to the fact that the neural network is sensitive to small perturbations, the training of the network is transferred to the image, which changes the problem that the quality of the steganographic scheme depends on the training of the extractor. Compared with existing steganographic schemes for training extractors, the training time of the proposed scheme is significantly reduced. At the same time, the sender and the receiver only need to share the architecture of the message extractor network and the random seed used to initialize its weight, and the actual weight of the network does not need to be shared, avoiding the transmission problem of the message extractor.

\subsubsection{Message extractor}
For message extractor selection, any neural network that can convert an image into a binary sequence can be used as a message extractor. When the image quality of the perturbed image is low a new random seed can be re-selected and a different random message extractor can be initialized, and this process can be repeated for optimization. Both the sender and receiver share the exact same extractor structure of the network architecture.

\subsubsection{Loss function}
The loss function of GASP during training is as follows, where D is the loss of the discriminator:
\begin{equation}
\label{eq:loss}
R_1(\mathbf{s})=\frac{l}{2}\parallel\nabla_{\mathbf{y}_1,...,\mathbf{y}_n}D(\mathbf{s})\parallel^2=\frac{l}{2}\sum_{\mathbf{y}_i}\parallel\nabla_{\mathbf{y}_i}D(\mathbf{s})\parallel^2
\end{equation}
The loss of steganography is as follows:
\begin{equation}
\label{eq:loss2}
\begin{aligned}
&\min_{\tilde{\boldsymbol{X}}}\underbrace{\langle\boldsymbol{M},\log F(\tilde{\boldsymbol{X}})\rangle+\langle(1-\boldsymbol{M}),\log(1-F(\tilde{\boldsymbol{X}}))\rangle}_{L_{BCE}},\\&\mathrm{s.t.}\|X-\tilde{X}\|_{\infty}\leq\epsilon\mathrm{~and~}0\leq\tilde{\boldsymbol{X}}\leq1,
\end{aligned}
\end{equation}
where $clip_{0}^{1}(x)=\max{(\min(x,1),0)}$, In the optimization process we use the unconstrained L-BFGS [ ] algorithm to optimize the objective about. This is because the L-BFGS algorithm tracks the second-order gradient statistic and enables faster optimization.

\subsection{Steganography Results and Discussion}
\subsubsection{Evaluation metrics}
In our evaluation of the proposed scheme, we assess its performance based on two key aspects: message extraction error rate and image quality. To do so, we employ the following three metrics:
The bit error rate: $\frac{\|M-\lfloor F(\tilde{X})\rceil\|_{1}}{HWD}$
Peak signal-to-noise ratio (PSNR):
\begin{equation}
\label{eq:MSE}
\mathbf{MSE}=\frac1{HW}\sum_{i=l}^H\sum_{i=l}^W\left[X_{i,j}-\widetilde{X}_{i,j}\right]^2
\end{equation}
\begin{equation}
\label{eq:PSNR}
\mathbf{PSNR}=20\mathrm{log}_{10}(max_X)-10\mathrm{log}_{10}(\mathbf{MSE})
\end{equation}
Peak Signal Noise Ratio (PSNR) is used to measure the gap between the original image and the steganographic image and is a commonly used measure of image quality. We use PSNR to determine the gap between the quality of the generated image and the real image and the quality of the image before and after steganography.
Structural Similarity Index(SSIM):
\begin{equation}
\label{eq:SSIM}
\mathbf{SSIM}=\frac{(2\mu_\mathbf{X}\mu_{\widetilde{\mathbf{X}}}+c_1)\big(2\sigma_\mathbf{X}\widetilde{\mathbf{X}}+c_2\big)\big(\sigma_\mathbf{X}^2+\sigma_{\widetilde{\mathbf{X}}}^2+c_2\big)}{\big(\mu_\mathbf{X}^2+\mu_{\widetilde{\mathbf{X}}}^2+c_1\big)\big(\sigma_\mathbf{X}^2+\sigma_{\widetilde{\mathbf{X}}}^2+c_2\big)}
\end{equation}
Where $c_1, c_2$ is a very small stability constant.SSIM is used to measure the similarity between the cover image X and the stega-image $\widetilde{X}$, which is different from PNSR in that PNSR considers pixel-level differences whereas SSIM considers structural differences.

\subsubsection{Settings}
During GASP training, we parameterize $f_{\theta}$ using a 3-layer multilayer perceptron, where each layer has 128 units. The hypernetwork $g_{\varphi}$ uses a 2-layer multilayer perceptron with vector dimensions of 64. The hypernetwork weights use an Adam optimizer with a learning rate of 1e-4 and a discriminator with a learning rate of 4e-4,$\beta_{1}=0.5$ and $\beta_{2}=0.999$. The batch size is 64 and the number of rounds is set to 300. the CelebA dataset size is 28000 and the weather data is from the ERA5 dataset.

For the steganography process, we design the message extractor based on the basic decoder of StegaGAN\cite{baluja2019hiding}. The extractor is a 4-layer convolutional neural network (including 3 intermediate layers and 1 output layer) with 128 hidden units in each intermediate layer, using a nonlinear Relu function to make the optimization more stable. The decoder takes an RGB image of H×W×3 as input and outputs a binary sequence of $\{0,1\}^{H\times W\times D}$. The parameters of the hidden writing process hypernetwork are set as follows: perturbation boundaries $\epsilon=0.3$, number of optimizations $n=100$, number of LBFGS iterations $k=10$, and the iteration can be terminated earlier if the extraction error rate is 0. Learning rate $\alpha=0.1$, when the PSNR value of the generated crypto-containing image is less than 20 choose a new random seed to re-optimize, during the experiment, the random seed is randomly selected from the pre-set ten values. The changed $\alpha$ into 0.5 or 0.05, and the steganographic capacity is from 1bpp to 4bpp to test the steganographic effect of the image under different parameters.

We evaluate the effectiveness of our scheme on two datasets, CelebA and ERA5. All training is performed on a 2.30GHz NVIDIA GeForce RTX 2070.

\subsection{Image Quality}
We use two metrics, PSNR and SSIM, to evaluate the quality of the images and to judge them from both qualitative and quantitative perspectives.

\subsubsection{Qualitative analysis}
\begin{figure}
\begin{center}
\begin{tabular}{c}
\includegraphics[height=5.5cm]{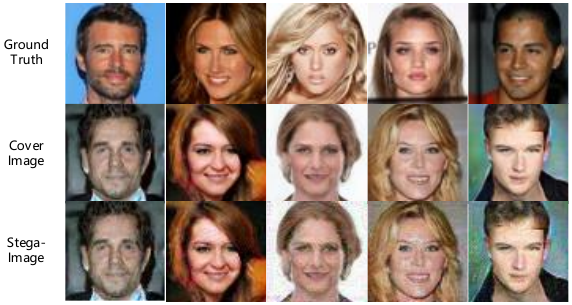}
\end{tabular}
\end{center}
\caption 
{ \label{fig:Qualitative Sampling}
Qualitative Sampling. } 
\end{figure}

Several experimental results are given in Figure \ref{fig:Qualitative Sampling}. There are a few examples of stega-images with noise points that appear pixelated. This may be due to the image being over-optimized, or to the use of bad random seeds. To improve the situation, the random network can be reinitialized to stop optimization earlier, or the hyperparameters of the steganography process can be slightly changed to produce higher-quality stega-images.

\begin{figure}
\begin{center}
\begin{tabular}{c}
\includegraphics[height=5.5cm]{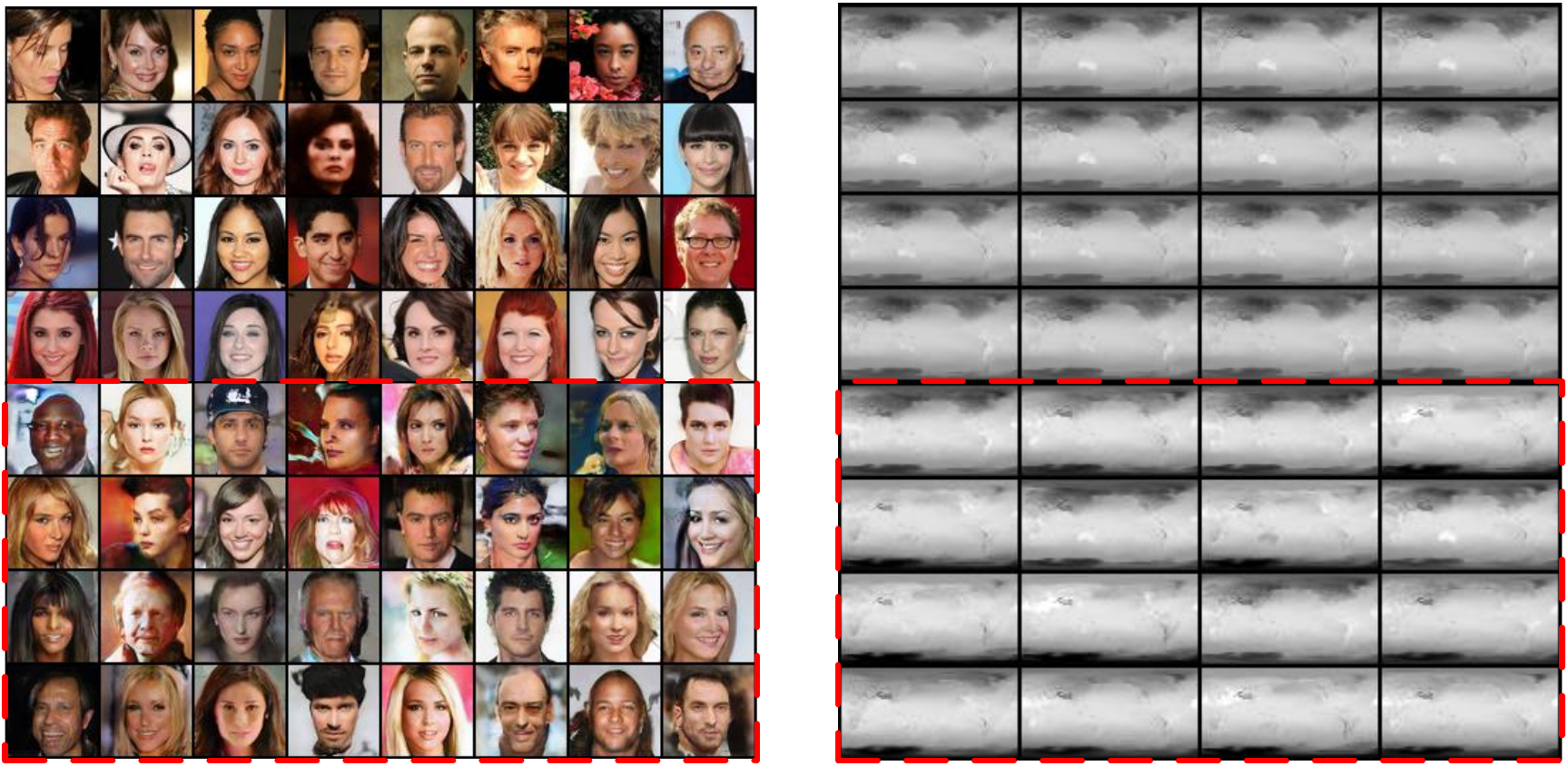}  
\\
(a) \hspace{5.1cm} (b)
\end{tabular}
\end{center}
\caption 
{ \label{fig:Sampling from different datasets}
Sampling from different datasets.The images above are ground trues and the red rectangles are the generated images: (a) CelebA and (b) ERA5. } 
\end{figure}

\begin{figure}
\begin{center}
\begin{tabular}{c}
\includegraphics[width=\textwidth]{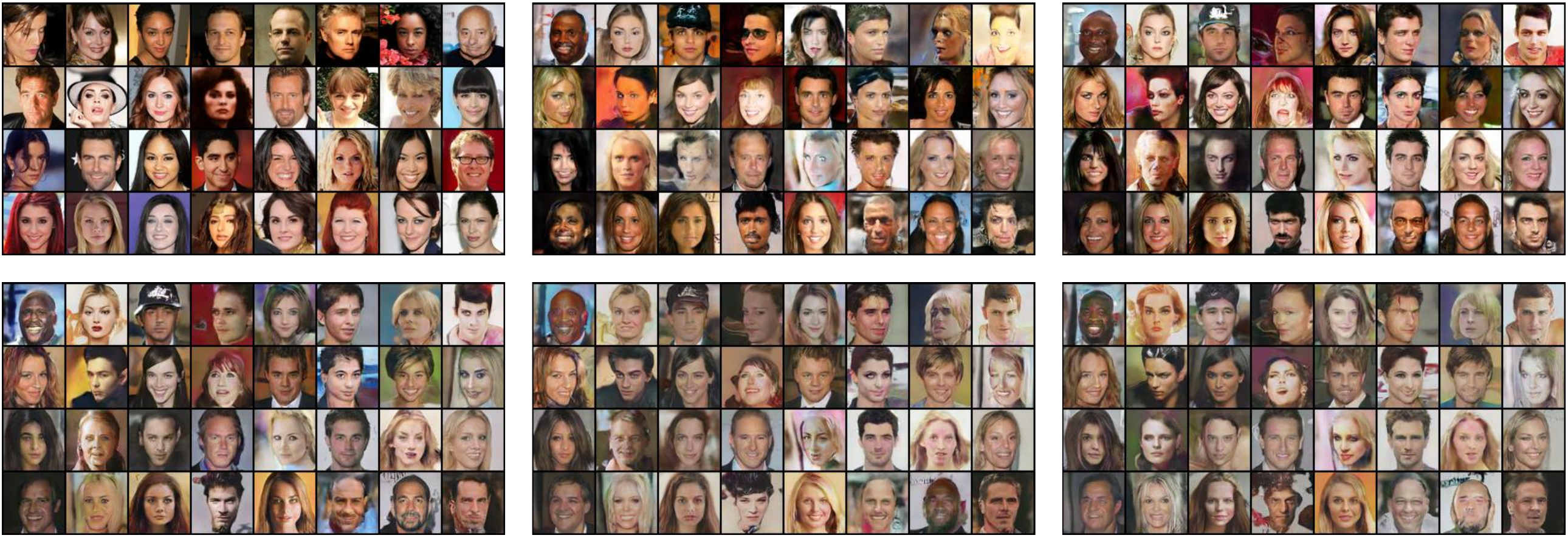}
\end{tabular}
\end{center}
\caption 
{ \label{fig:Images with different number of iterations}
Images with different number of iterations. } 
\end{figure}

Figure \ref{fig:Sampling from different datasets} gives the generation results under the CelebA and ERA5 datasets, which extend the variety of steganographic vectors by representing the various data as functions. Figure \ref{fig:Images with different number of iterations} gives the quality of the generated images for different numbers of optimizations of the model, the first one is the real image sampled from the dataset and the last five are the sampled images with 50, 100, 150, 200 and 250 optimizations. As the number of optimizations increases the quality of the image increases significantly, but when the optimization is gradually increased, the color richness of the image decreases significantly, which may be due to the occurrence of overfitting. The number of model optimizations is not the more the better, according to the actual situation to determine the flexibility, in the optimization of the number of times for about 100 times when the image generation effect is the best.

\begin{figure}
\begin{center}
\begin{tabular}{c}
\includegraphics[height=5.5cm]{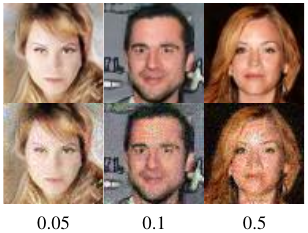}
\end{tabular}
\end{center}
\caption 
{ \label{fig:Sampling in different learning rates}
Sampling in different learning rates. } 
\end{figure}

Meanwhile, to test the gradient descent and optimization process, we choose different learning rates for the experiments, and Figure \ref{fig:Sampling in different learning rates} shows the generated stega-images under different learning rates.

\begin{figure}
\begin{center}
\begin{tabular}{c}
\includegraphics[height=5.5cm]{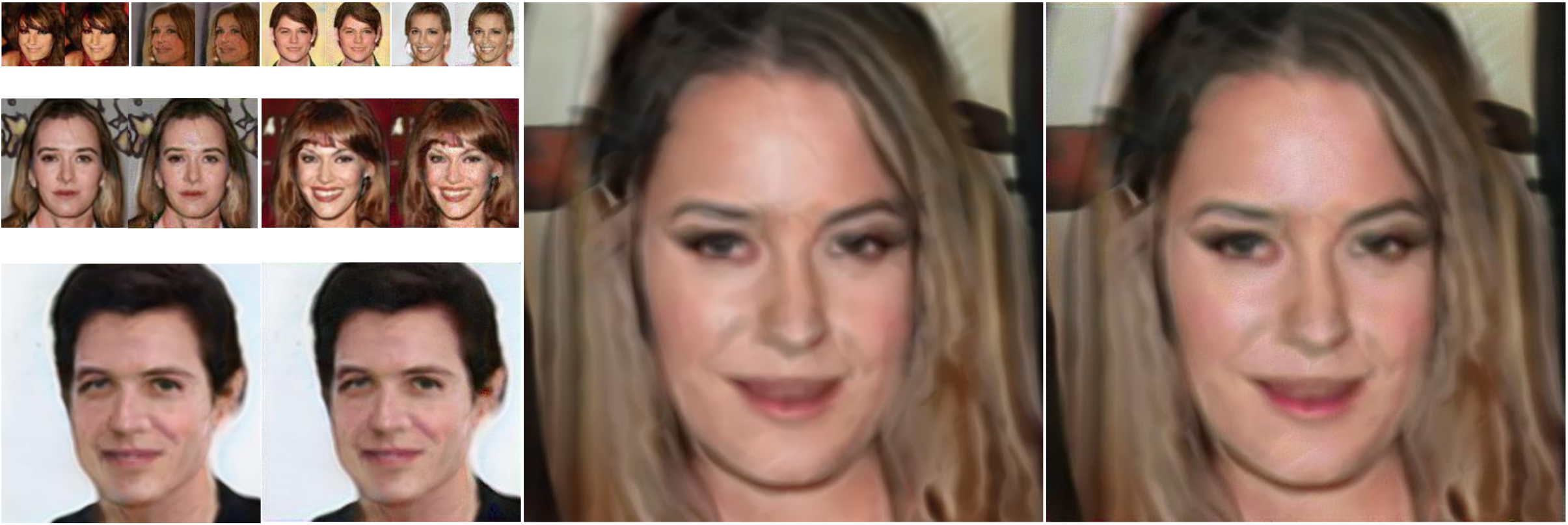}
\end{tabular}
\end{center}
\caption 
{ \label{fig:Stega-image in different resolutions}
Stega-image in different resolutions. } 
\end{figure}

Figure \ref{fig:Stega-image in different resolutions} shows the super-resolution sampled image and the image after steganography under the training model of 64×64 dimensions, the cover image can be sampled at any resolution from a model trained at a fixed resolution, and the experiment proves that the steganography process applies to all kinds of resolution images.

\subsection{Quantitative analysis}
\begin{figure}
\begin{center}
\begin{tabular}{c}
\includegraphics[height=5.5cm]{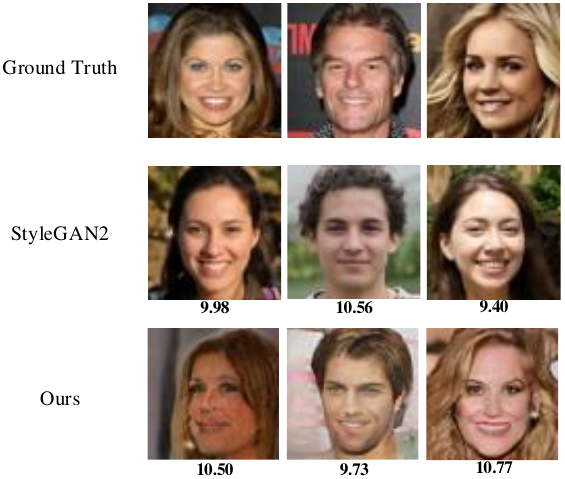}
\end{tabular}
\end{center}
\caption 
{ \label{fig:To compare PSNR with images generated by StyleGAN2}
To compare PSNR with images generated by StyleGAN2. } 
\end{figure}

We evaluate the PSNR and SSIM effects in three datasets of CelebA, Div2k, and MS-coco respectively. The comparison data are from the literature\cite{kishore2022fixed}. PSNR-1 is the stega-image compared with the cover image, and PSNR-2 is the cover image compared with the real image. The experimental results are shown in Table \ref{tab:Image Quality Under Different Datasets}. Due to the limitation of the function generation model itself, the quality of the generated images has not yet Due to the limitations of the function generation model itself, the quality of the generated images has not reached a satisfactory level, but as far as the steganography effect is concerned, the steganographic image of GINR-Stega has an obvious advantage in image similarity with the cover image. To further evaluate the generation effect of GASP, we compare the PSNR value with the image generated by StyleGAN2, and the results are shown in Figure \ref{fig:To compare PSNR with images generated by StyleGAN2}.

\begin{table}[ht]
\caption{Image Quality Under Different Datasets.} 
\label{tab:Image Quality Under Different Datasets}
\begin{center}       
\begin{tabular}{|c|c|c|c|}
\hline
\rule[-1ex]{0pt}{3.5ex}  Datasets&PSNR-1&SSIM&PSNR-2  \\
\hline\hline
\rule[-1ex]{0pt}{3.5ex}  Div2k & 38.41 & 0.93 & 0  \\
\hline
\rule[-1ex]{0pt}{3.5ex}  MS-coco & 30.76 & 0.91 & 0   \\
\hline 
\end{tabular}
\end{center}
\end{table} 

\begin{figure}
\begin{center}
\begin{tabular}{c}
\includegraphics[width=\linewidth]{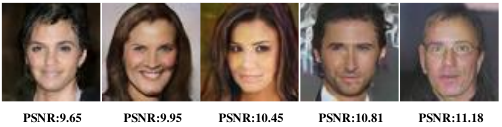}
\end{tabular}
\end{center}
\caption 
{ \label{fig:The quality of cover image under different training rounds}
The quality of cover image under different training rounds. } 
\end{figure}

Figure \ref{fig:The quality of cover image under different training rounds} gives the quality of the images sampled under the model with training rounds of 50, 100, 150, 200, and 250, respectively, and it can be seen that the quality of the images improves significantly as the dataset grows larger and the number of training rounds increases, although it has not yet reached the ideal state for the time being due to the limitations of the generative model itself, it can be demonstrated that there is a great potential for using functions for data representation.

\subsection{Capacity}
\begin{figure}
\begin{center}
\begin{tabular}{c}
\includegraphics[height=5.5cm]{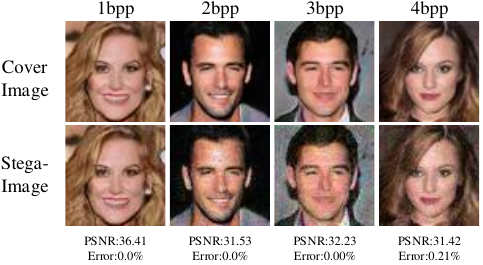}
\end{tabular}
\end{center}
\caption 
{ \label{fig:The Quality in Different Bits Per Pixel}
The Quality in Different Bits Per Pixel. } 
\end{figure}

Figure \ref{fig:The Quality in Different Bits Per Pixel} demonstrates the various metrics of the image in different hiding capacities, even though the message of 3bpp is hidden, the steganographic image still looks not much different from the cover image, but as the steganographic capacity increases, the image quality tends to decrease and the pixelation becomes more obvious.

To evaluate our scheme, we perform a quantitative comparison with Stegano-GAN, and two variants of FNNS, with comparison data from Kishore\cite{kishore2022fixed}. Table \ref{tab:Quantitative comparison} demonstrates the BER of messages at different steganographic capacities, and 100\% message extraction accuracy can still be achieved at a steganographic capacity of 3bpp. The values of PSNR and SSIM can be seen from Table \ref{tab:The quality of images in different bpps} and GINR-Stega outperforms the other three schemes in terms of image quality and structural similarity. Due to the limitations of the function generation model and the size of the dataset, there is still a gap between the quality of our images and the real images to the naked eye, but its demonstrated image representation and generation capability have proved its great potential.

\begin{table}[ht]
\caption{Quantitative comparison.} 
\label{tab:Quantitative comparison}
\begin{center}       
\begin{tabular}{|c|c|c|c|c|} 
\hline
\rule[-1ex]{0pt}{3.5ex}  Methods     & 1bpp     & 2bpp     & 3bpp      & 4bpp  \\
\hline\hline
\rule[-1ex]{0pt}{3.5ex}  StegaGAN    & 3.94     & 7.36     & 8.84      & 10.00  \\
\hline
\rule[-1ex]{0pt}{3.5ex}  FNNS-D      & 0.00     & 0.00     & 0.00      & 3.17   \\
\hline
\rule[-1ex]{0pt}{3.5ex}  FNNS-DE     & 0.00     & 0.00     & 0.00      & 2.58   \\
\hline
\rule[-1ex]{0pt}{3.5ex}  Ours        & 0.00     & 0.00     & 0.00      & 0.19  \\
\hline 
\end{tabular}
\end{center}
\end{table} 

\begin{table}[ht]
\caption{The quality of images in different bpps.} 
\label{tab:The quality of images in different bpps}
\begin{center}       
\begin{tabular}{|c|c|c|c|c|c|c|c|c|} 
\hline
\rule[-1ex]{0pt}{3.5ex}  Methods     & 1bpp     & 2bpp     & 3bpp      & 4bpp    & 1bpp     & 2bpp     & 3bpp      & 4bpp \\
\hline\hline
\rule[-1ex]{0pt}{3.5ex} StegaGAN       & 25.98          & 25.53          & 25.70          & 25.08          & 0.85          & 0.86          & 0.85          & 0.82  \\
\hline
\rule[-1ex]{0pt}{3.5ex}  FNNS-D         & 36.06          & 34.43          & 30.05          & 33.92          & 0.87          & 0.86          & 0.71          & 0.84   \\
\hline
\rule[-1ex]{0pt}{3.5ex}  FNNS-DE        & 21.16          & 20.85          & 20.67          & 21.03          & 0.71          & 0.68          & 0.63          & 0.64   \\
\hline
\rule[-1ex]{0pt}{3.5ex}  Ours        & 34.50          & 28.37          & 29.64          & 27.14          & 0.95          & 0.82          & 0.87          & 0.83  \\
\hline 
\end{tabular}
\end{center}
\end{table} 

\subsection{Efficiency}
\begin{figure}
\begin{center}
\begin{tabular}{c}
\includegraphics[width=\linewidth]{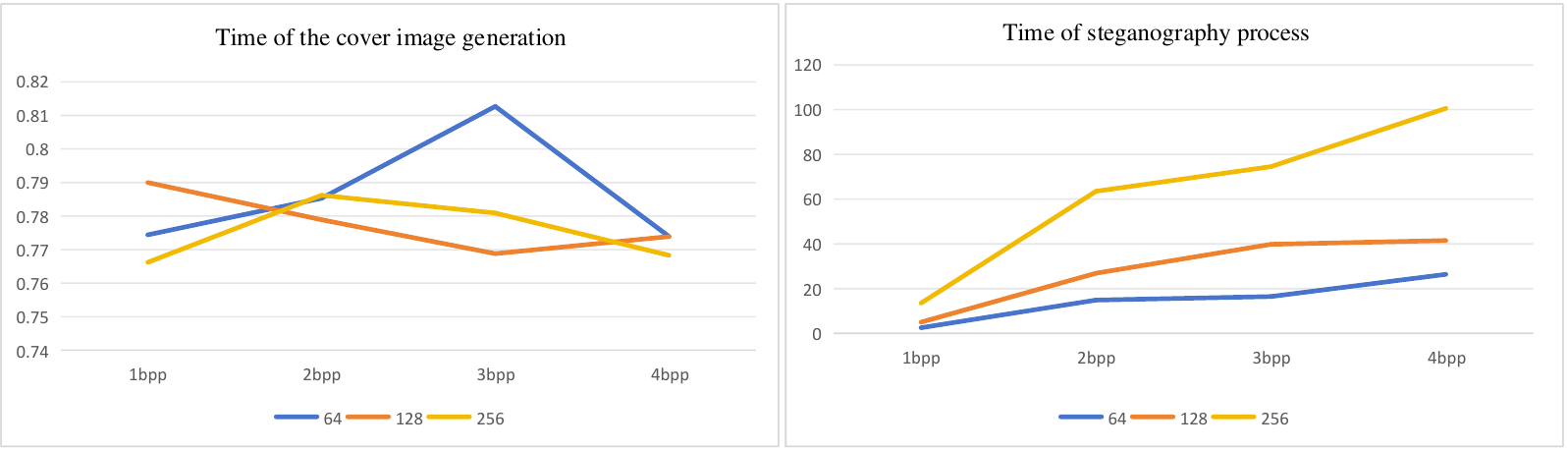}
\end{tabular}
\end{center}
\caption 
{ \label{fig:optimization time}
Image optimization time under different dimensions and hidden capacities. } 
\end{figure}

It takes 75 hours to train GASP in the set experimental parameters, although the pre-training time is long, the trained model can be applied to images with different resolutions and the trained model takes less than 1 second to react to a given cover image and message, 1bpp takes only 3 seconds and 4bpp takes about 27 seconds to complete the optimization. Table \ref{tab:Optimization Time} shows the time used to generate the stega-image at 64×64 resolution for the CelebA dataset as an example, with an image size of and with different amounts of bits embedded. Where t denotes the original cover image generation time and T denotes the steganography optimization time. We also try to vary the image size and the hidden capacity for exploration, and experiments prove that the image size and the encoding time are approximately linearly related. Figure \ref{fig:optimization time} demonstrates the image optimization time versus image size and hiding capacity. It can be seen that the cover image generation time is almost the same regardless of the resolution, and the steganography optimization time is approximately linearly related to the hiding capacity and the image dimension.

\begin{table}[ht]
\caption{Optimization Time.} 
\label{tab:Optimization Time}
\begin{center}       
\begin{tabular}{|c|c|c|c|c|} 
\hline
\rule[-1ex]{0pt}{3.5ex}  Time     & 1bpp   & 2bpp    & 3bpp    & 4bpp  \\
\hline\hline
\rule[-1ex]{0pt}{3.5ex}  t & 0.7744 & 0.7853  & 0.8127  & 0.7739  \\
\hline
\rule[-1ex]{0pt}{3.5ex}  T & 2.5067 & 14.8670 & 16.4525 & 26.3781   \\
\hline 
\end{tabular}
\end{center}
\end{table} 

\subsection{Security analysis}
Statistical analysis is a steganography tool. It uses statistical methods to detect stega-images.We use the Stegexpose method proposed by Boehmm et al.\cite{boehmm2014stegexpose} to detect stega-images on the CelebA dataset, and the results are shown in Table \ref{tab:detection}. Even though a portion of the images will be detected and discovered by the detection, the detection rate is low compared to the traditional steganographic schemes, with an accuracy rate of more than 75\%\cite{tancik2020stegastamp}.

\begin{table}[ht]
\caption{Detection Result using Stegexpose on the CelebA Dataset.} 
\label{tab:detection}
\begin{center}       
\begin{tabular}{|c|c|c|c|c|} 
\hline
\rule[-1ex]{0pt}{3.5ex}  Methods     & 1bpp   & 2bpp    & 3bpp    & 4bpp  \\
\hline\hline
\rule[-1ex]{0pt}{3.5ex}  FNNS-D      & 17         & 13          & 15          & 8   \\
\hline
\rule[-1ex]{0pt}{3.5ex} FNNS-DE      & 2          & 3           & 0           & 0   \\
\hline 
\rule[-1ex]{0pt}{3.5ex} Ours      & 7         & 16           & 26           & 33   \\
\hline 
\end{tabular}
\end{center}
\end{table} 

\section{Conclusion}
In this paper, we propose for function generative image steganography based on implicit
neural expression for the first time. Using the concept of implicit neural representation, the cover image is generated using a generative adversarial stochastic process, and a fixed neural network is used to hide information on the cover image, experiments prove the feasibility of our scheme. Our scheme can achieve a very low error rate while making visually insignificant changes and can achieve a 100\% extraction rate at an embedding capacity of 3bpp. At the same time, the fixed extractor format allows us to eliminate the need to spend a large amount of resources on extractor training, shifting the focus to the training process of the image, which greatly reduces the training cost. The current message extractor used in our scheme is also applicable to explicit images, and in the future, we will change the message extractor to make it work directly on point cloud data to adapt to all kinds of multimedia carriers.

\subsection*{Disclosures}
We declare that they have no conflicts of interest to disclose.

\subsection* {Code and Data Availability} 
The archived version of the code described in this manuscript can be freely accessed through GitHub [https://github.com/twinlj77/GFNNS].

\subsection* {Acknowledgments}
This work was part of the Science and Technology Innovation Team Innovative Research Program(ZZKY20222102) and was supported in part by the National Natural Science Foundation of China under Grant 62102450, Grant 61872384, and Grant 62272478.


\bibliography{report}   

\begin{thebibliography}{10}

\bibitem{xu2014data}
Dawen~Xu 0001, Rangding Wang, and Yun~Q. Shi.
\newblock Data hiding in encrypted h.264/avc video streams by codeword substitution.
\newblock {\em IEEE Trans. Information Forensics and Security}, 9:596--606, 2014.

\bibitem{baluja2019hiding}
Shumeet Baluja.
\newblock Hiding images within images.
\newblock {\em IEEE transactions on pattern analysis and machine intelligence}, 42(7):1685--1697, 2019.

\bibitem{boehmm2014stegexpose}
E~Boehmm.
\newblock Stegexpose: a tool for detecting lsb steganography (2014), 2014.

\bibitem{borges2008robust}
Paulo Vinicius~Koerich Borges, Joceli Mayer, and Ebroul Izquierdo.
\newblock Robust and transparent color modulation for text data hiding.
\newblock {\em IEEE Trans. Multimedia}, 10:1479--1489, 2008.

\bibitem{chen2023marknerf}
Lifeng Chen, Jia Liu, Yan Ke, Wenquan Sun, Weina Dong, and Xiaozhong Pan.
\newblock Marknerf: Watermarking for neural radiance field.
\newblock {\em arXiv preprint arXiv:2309.11747}, 2023.

\bibitem{chen2019learning}
Zhiqin Chen and Hao Zhang.
\newblock Learning implicit fields for generative shape modeling.
\newblock In {\em Proceedings of the IEEE/CVF conference on computer vision and pattern recognition}, pages 5939--5948, 2019.

\bibitem{dong2023steganography}
Weina Dong, Jia Liu, Yan Ke, Lifeng Chen, Wenquan Sun, and Xiaozhong Pan.
\newblock Steganography for neural radiance fields by backdooring.
\newblock {\em arXiv preprint arXiv:2309.10503}, 2023.

\bibitem{dupont2022generative}
Emilien Dupont, Yee~Whye Teh, and Arnaud Doucet.
\newblock Generative models as distributions of functions, 2021.

\bibitem{fridrich2009steganography}
Jessica Fridrich.
\newblock {\em Steganography in Digital Media}.
\newblock Cambridge University Press, 2009.

\bibitem{goodfellow2014generative}
Ian Goodfellow, Jean Pouget-Abadie, Mehdi Mirza, Bing Xu, David Warde-Farley, Sherjil Ozair, Aaron Courville, and Yoshua Bengio.
\newblock Generative adversarial nets.
\newblock In Z.~Ghahramani, M.~Welling, C.~Cortes, N.~Lawrence, and K.Q. Weinberger, editors, {\em Advances in Neural Information Processing Systems}, volume~27. Curran Associates, Inc., 2014.

\bibitem{goodfellow2020generative}
Ian Goodfellow, Jean Pouget-Abadie, Mehdi Mirza, Bing Xu, David Warde-Farley, Sherjil Ozair, Aaron Courville, and Yoshua Bengio.
\newblock Generative adversarial networks.
\newblock {\em Communications of the ACM}, 63(11):139--144, 2020.

\bibitem{guan2022deepmih}
Zhenyu Guan, Junpeng Jing, Xin Deng, Mai Xu, Lai Jiang, Zhou Zhang, and Yipeng Li.
\newblock Deepmih: Deep invertible network for multiple image hiding.
\newblock {\em IEEE Transactions on Pattern Analysis and Machine Intelligence}, 45(1):372--390, 2022.

\bibitem{ha2016generating}
David Ha.
\newblock Generating large images from latent vectors.
\newblock 2016.

\bibitem{ha2016hypernetworks}
David Ha, Andrew Dai, and Quoc~V Le.
\newblock Hypernetworks.
\newblock {\em arXiv preprint arXiv:1609.09106}, 2016.

\bibitem{han2023deep}
Gyojin Han, Dong-Jae Lee, Jiwan Hur, Jaehyun Choi, and Junmo Kim.
\newblock Deep cross-modal steganography using neural representations.
\newblock In {\em 2023 IEEE International Conference on Image Processing (ICIP)}, pages 1205--1209. IEEE, 2023.

\bibitem{ho2020denoising}
Jonathan Ho, Ajay Jain, and Pieter Abbeel.
\newblock Denoising diffusion probabilistic models.
\newblock {\em Advances in neural information processing systems}, 33:6840--6851, 2020.

\bibitem{hu2018novel}
Donghui Hu, Liang Wang, Wenjie Jiang, Shuli Zheng, and Bin Li.
\newblock A novel image steganography method via deep convolutional generative adversarial networks.
\newblock {\em IEEE access}, 6:38303--38314, 2018.

\bibitem{jing2021hinet}
Junpeng Jing, Xin Deng, Mai Xu, Jianyi Wang, and Zhenyu Guan.
\newblock Hinet: Deep image hiding by invertible network.
\newblock In {\em Proceedings of the IEEE/CVF international conference on computer vision}, pages 4733--4742, 2021.

\bibitem{karras2022elucidating}
Tero Karras, Miika Aittala, Timo Aila, and Samuli Laine.
\newblock Elucidating the design space of diffusion-based generative models.
\newblock {\em Advances in Neural Information Processing Systems}, 35:26565--26577, 2022.

\bibitem{ke2019generative}
Yan Ke, Min-qing Zhang, Jia Liu, Ting-ting Su, and Xiao-yuan Yang.
\newblock Generative steganography with kerckhoffs’ principle.
\newblock {\em Multimedia Tools and Applications}, 78(10):13805--13818, 2019.

\bibitem{kim2023diffusion}
Daegyu Kim, Chaehun Shin, Jooyoung Choi, Dahuin Jung, and Sungroh Yoon.
\newblock Diffusion-stego: Training-free diffusion generative steganography via message projection.
\newblock {\em arXiv preprint arXiv:2305.18726}, 2023.

\bibitem{kingma2013auto}
Diederik~P Kingma and Max Welling.
\newblock Auto-encoding variational bayes.
\newblock {\em arXiv preprint arXiv:1312.6114}, 2013.

\bibitem{kingma2018glow}
Durk~P Kingma and Prafulla Dhariwal.
\newblock Glow: Generative flow with invertible 1x1 convolutions.
\newblock {\em Advances in neural information processing systems}, 31, 2018.

\bibitem{kishore2022fixed}
Varsha Kishore, Xiangyu Chen, Yan Wang, Boyi Li, and Kilian~Q Weinberger.
\newblock Fixed neural network steganography: Train the images, not the network.
\newblock In {\em International Conference on Learning Representations}, 2021.

\bibitem{li2023steganerf}
Chenxin Li, Brandon~Y Feng, Zhiwen Fan, Panwang Pan, and Zhangyang Wang.
\newblock Steganerf: Embedding invisible information within neural radiance fields.
\newblock In {\em Proceedings of the IEEE/CVF International Conference on Computer Vision}, pages 441--453, 2023.

\bibitem{li2023towards}
Guobiao Li, Sheng Li, Meiling Li, Zhenxing Qian, and Xinpeng Zhang.
\newblock Towards deep network steganography: From networks to networks.
\newblock {\em arXiv preprint arXiv:2307.03444}, 2023.

\bibitem{li2023steganography}
Guobiao Li, Sheng Li, Meiling Li, Xinpeng Zhang, and Zhenxing Qian.
\newblock Steganography of steganographic networks.
\newblock In {\em Proceedings of the AAAI Conference on Artificial Intelligence}, volume~37, pages 5178--5186, 2023.

\bibitem{liu2023hiding}
Jia Liu, Peng Luo, and Yan Ke.
\newblock Hiding functions within functions: Steganography by implicit neural representations.
\newblock {\em arXiv preprint arXiv:2312.04743}, 2023.

\bibitem{liu2018digital}
Jia Liu, Tanping Zhou, Zhuo Zhang, Yan Ke, Yu~Lei, and Minqing Zhang.
\newblock Digital cardan grille: A modern approach for information hiding.
\newblock In {\em Proceedings of the 2018 2nd International Conference on Computer Science and Artificial Intelligence}, pages 44--446, 2018.

\bibitem{luo2023copyrnerf}
Ziyuan Luo, Qing Guo, Ka~Chun Cheung, Simon See, and Renjie Wan.
\newblock Copyrnerf: Protecting the copyright of neural radiance fields.
\newblock In {\em Proceedings of the IEEE/CVF International Conference on Computer Vision}, pages 22401--22411, 2023.

\bibitem{2018Which}
Lars Mescheder, Andreas Geiger, and Sebastian Nowozin.
\newblock Which training methods for gans do actually converge?
\newblock 2018.

\bibitem{mescheder2020stability}
Lars~Morten Mescheder.
\newblock {\em Stability and Expressiveness of Deep Generative Models}.
\newblock PhD thesis, Universit{\"a}t T{\"u}bingen, 2020.

\bibitem{mielikainen2006lsb}
Jarno Mielikainen.
\newblock Lsb matching revisited.
\newblock {\em IEEE signal processing letters}, 13(5):285--287, 2006.

\bibitem{otori2007data}
Hirofumi Otori and Shigeru Kuriyama.
\newblock Data-embeddable texture synthesis.
\newblock In {\em International Symposium on Smart Graphics}, pages 146--157. Springer, 2007.

\bibitem{Park_2019_CVPR}
Jeong~Joon Park, Peter Florence, Julian Straub, Richard Newcombe, and Steven Lovegrove.
\newblock Deepsdf: Learning continuous signed distance functions for shape representation.
\newblock In {\em Proceedings of the IEEE/CVF conference on computer vision and pattern recognition}, pages 165--174, 2019.

\bibitem{shi2018ssgan}
Haichao Shi, Jing Dong, Wei Wang, Yinlong Qian, and Xiaoyu Zhang.
\newblock Ssgan: Secure steganography based on generative adversarial networks.
\newblock In {\em Advances in Multimedia Information Processing--PCM 2017: 18th Pacific-Rim Conference on Multimedia, Harbin, China, September 28-29, 2017, Revised Selected Papers, Part I 18}, pages 534--544. Springer, 2018.

\bibitem{sitzmann2020implicit}
Vincent Sitzmann, Julien Martel, Alexander Bergman, David Lindell, and Gordon Wetzstein.
\newblock Implicit neural representations with periodic activation functions.
\newblock {\em Advances in neural information processing systems}, 33:7462--7473, 2020.

\bibitem{sohl2015deep}
Jascha Sohl-Dickstein, Eric Weiss, Niru Maheswaranathan, and Surya Ganguli.
\newblock Deep unsupervised learning using nonequilibrium thermodynamics.
\newblock In {\em International conference on machine learning}, pages 2256--2265. PMLR, 2015.

\bibitem{tancik2020stegastamp}
Matthew Tancik, Ben Mildenhall, and Ren Ng.
\newblock Stegastamp: Invisible hyperlinks in physical photographs.
\newblock In {\em Proceedings of the IEEE/CVF conference on computer vision and pattern recognition}, pages 2117--2126, 2020.

\bibitem{tao2018towards}
Jinyuan Tao, Sheng Li, Xinpeng Zhang, and Zichi Wang.
\newblock Towards robust image steganography.
\newblock {\em IEEE Trans. Circuits Syst. Video Techn.}, 29:594--600, 2019.

\bibitem{van1994digital}
Ron~G Van~Schyndel, Andrew~Z Tirkel, and Charles~F Osborne.
\newblock A digital watermark.
\newblock In {\em Proceedings of 1st international conference on image processing}, volume~2, pages 86--90. IEEE, 1994.

\bibitem{volkhonskiy2020steganographic}
Denis Volkhonskiy, Ivan Nazarov, and Evgeny Burnaev.
\newblock Steganographic generative adversarial networks.
\newblock In {\em Twelfth international conference on machine vision (ICMV 2019)}, volume 11433, pages 991--1005. SPIE, 2020.

\bibitem{wei2023generative}
Ping Wei, Qing Zhou, Zichi Wang, Zhenxing Qian, Xinpeng Zhang, and Sheng Li.
\newblock Generative steganography diffusion.
\newblock {\em arXiv preprint arXiv:2305.03472}, 2023.

\bibitem{tang2017automatic}
Tang Weixuan, Tan Shunquan, Li~Bin, and Huang Jiwu.
\newblock Automatic steganographic distortion learning using a generative adversarial network.
\newblock {\em IEEE Signal Processing Letters}, 24:1547--1551, 2017.

\bibitem{wu2019pointconv}
Wenxuan Wu, Zhongang Qi, and Li~Fuxin.
\newblock Pointconv: Deep convolutional networks on 3d point clouds.
\newblock In {\em Proceedings of the IEEE/CVF Conference on computer vision and pattern recognition}, pages 9621--9630, 2019.

\bibitem{xu2023pfgm++}
Yilun Xu, Ziming Liu, Yonglong Tian, Shangyuan Tong, Max Tegmark, and Tommi Jaakkola.
\newblock Pfgm++: Unlocking the potential of physics-inspired generative models.
\newblock In {\em International Conference on Machine Learning}, pages 38566--38591. PMLR, 2023.

\bibitem{yang2018spatial}
Jianhua Yang, Kai Liu, Xiangui Kang, Edward~K Wong, and Yun-Qing Shi.
\newblock Spatial image steganography based on generative adversarial network.
\newblock {\em arXiv preprint arXiv:1804.07939}, 2018.

\bibitem{yang2023flexible}
Seoyun Yang, Sojeong Song, Chang~D Yoo, and Junmo Kim.
\newblock Flexible cross-modal steganography via implicit representations.
\newblock {\em arXiv preprint arXiv:2312.05496}, 2023.

\bibitem{yi2019ahcm}
Xiaowei Yi, Kun Yang, Xianfeng Zhao, Yuntao Wang, and Haibo Yu.
\newblock Ahcm: Adaptive huffman code mapping for audio steganography based on psychoacoustic model.
\newblock {\em IEEE Trans. Information Forensics and Security}, 14:2217--2231, 2019.

\bibitem{zhang2020udh}
Chaoning Zhang, Philipp Benz, Adil Karjauv, Geng Sun, and In~So Kweon.
\newblock Udh: Universal deep hiding for steganography, watermarking, and light field messaging.
\newblock {\em Advances in Neural Information Processing Systems}, 33:10223--10234, 2020.

\bibitem{zhang2019steganogan}
Kevin~Alex Zhang, Alfredo Cuesta-Infante, Lei Xu, and Kalyan Veeramachaneni.
\newblock Steganogan: High capacity image steganography with gans.
\newblock {\em CoRR}, abs/1901.03892, 2019.

\bibitem{zhu2018hidden}
Jiren Zhu, Russell Kaplan, Justin Johnson, and Li~Fei-Fei.
\newblock Hidden: Hiding data with deep networks.
\newblock In {\em Proceedings of the European conference on computer vision (ECCV)}, pages 657--672, 2018.

\end{thebibliography}
\bibliographystyle{spiejour}   


\vspace{2ex}\noindent\textbf{Yangjie Zhong} is currently a graduate student in the Computer Technology program with a focus on deep learning and steganography.

\vspace{1ex}
\noindent Biographies and photographs of the other authors are not available.

\listoffigures
\listoftables

\end{spacing}
\end{document}